# ENERGY FILTERING WITH X-RAY LENSES: OPTIMIZATION FOR PHOTON-COUNTING MAMMOGRAPHY

Erik Fredenberg [a,*], Björn Cederström [a], Mats Danielsson [a]
[a]Department of Physics, Royal Institute of Technology, AlbaNova, SE-106 91 Stockholm, Sweden

**Chromatic properties of the multi-prism and prism-array x-ray lenses (MPL and PAL) can potentially be utilized for efficient energy filtering and dose reduction in mammography. The line-shaped foci of the lenses are optimal for coupling to photon-counting silicon strip detectors in a scanning system. A theoretical model was developed and used to investigate the benefit of two lenses compared to an absorption-filtered reference system. The dose reduction of the MPL filter was $\sim 15\%$ compared to the reference system at matching scan time, and the spatial resolution was higher. The dose of the PAL-filtered system was found to be $\sim 20\%$ lower than for the reference system at equal scan time and resolution, and only $\sim 20\%$ higher than for a monochromatic beam. An investigation of some practical issues remains, including the feasibility of brilliant-enough x-ray sources and manufacturing of a polymer PAL.**

## INTRODUCTION

In medical x-ray imaging, there exists an optimum in incident photon energy to maximize contrast, and minimize quantum noise and dose.[1] This issue is particularly important to consider in mammography because many women are exposed to radiation within screening programs. Optimization of the x-ray spectrum with respect to energy is known as spectral shaping or energy filtering, and absorption filtering has been the dominant means to achieve this since the advent of medical x-ray imaging.[2–4] In principle, the spectrum can be made arbitrarily narrow by heavy K-edge filtration and a limited acceleration voltage, but only at the cost of a reduced flux.

Spectral shaping beyond the practical limit of absorption filtering can be achieved with synchrotrons, and some other exotic x-ray sources, which are capable of producing nearly monochromatic beams.[5,6] The high complexity and cost of such sources limit the feasibility for routine clinical x-ray imaging, however, and it is desirable to instead apply efficient filtering to conventional x-ray tubes. In that case, the relatively low photon yield becomes a problem because a large part of the spectrum is removed. Filtering with mosaic crystals with small imperfections in the crystal structure have been proposed as a suitable compromise between reflectivity and energy resolution.[7] A second option along the same avenue is energy filtering with curved crystals, which have a large solid angle of acceptance.[8]

A different approach is to use refractive, chromatic x-ray lenses.[9,10] These are inserted inline and therefore do not change the imaging geometry to the same extent as crystals do, and the provided line foci are ideal for scanning mammography systems.[4,11] Refractive lenses are manufactured for hard x-rays by dividing the weak refractive effect over a large number of surfaces, which is implemented in the multi-prism lens (MPL) with two rows of prisms at an angle in relation to the optical axis.[12] Advantages of the MPL in terms of small-scale energy filtering include a tunable focal length and relatively simple manufacturing, but the Gaussian transmission profile limits the usable aperture.

One way to increase lens transmission is to remove lens material corresponding to a phase shift of integer steps of $2\pi$.[13,14] This strategy is implemented in the prism-array lens (PAL), where each prism in the MPL is exchanged for a column of smaller prisms. Compared to an MPL, tunability is sacrificed, the complexity is higher, and the energy resolution of a PAL is lower, but the aperture can be made substantially larger, the lens is shorter, and the complexity is still lower than would be the case for a Fresnel lens.

We have developed a computer model of a lens-filtered mammography system, which is based on previous experimental and theoretical findings.[9,10] The model is employed to compare MPL, PAL, and absorption filters in terms of dose and signal-to-noise ratio with spatial resolution, exposure time, and imaging geometry as constraints.

## MATERIALS AND METHODS

Figure 1 compares the MPL and PAL. The projection of the MPL approximates a parabola with straight line segments,[12] whereas the PAL projection approximates a Fresnel lens, superimposed on a linear profile.[14] The focal lengths of the two lenses are the same if the prism angle ($\theta$) and the columnar displacement ($d$) are equal, and the base of each small prism corresponds to an integer number of $2\pi$ phase shifts.

---

*fberg@mi.physics.kth.se

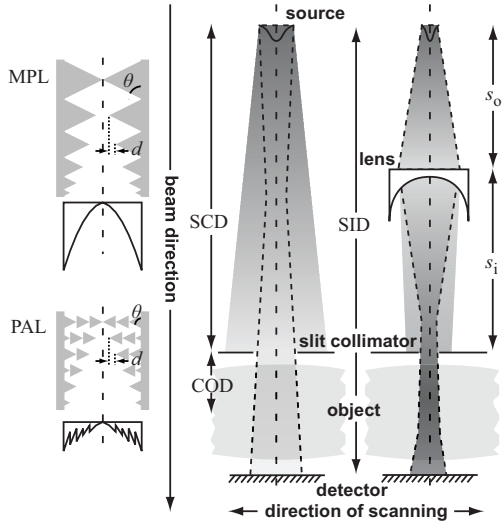

Figure 1. **Left:** Comparison of the MPL and PAL. **Right:** The reference and lens geometries.

Both lenses are planar, and therefore focuses radiation into a line focus. Further, the lenses are chromatic with focal lengths $F_{\mathrm{MPL}} \propto E^{2}$ [15] and $F_{\mathrm{PAL}} \propto E$,[10] where $E$ is the photon energy. Therefore, a slit placed in the image plane of a particular x-ray energy will transmit radiation of that energy, whereas other parts of a polychromatic incident beam are out of focus and preferentially blocked. We refer to a lens-slit combination as a filter. Due to the difference in focal length energy dependence, the energy resolution of a PAL filter is approximately 1.5 times lower than that of an MPL filter for equally transmitting lenses.[10] The filtered beam is fan shaped and matches well the photon-counting silicon strip detectors that are found in some scanning multi-slit mammography systems.[4, 11]

The geometries for comparison of lens- and absorption-filtering are outlined in Fig. 1, and are referred to as the lens and reference geometries. A full system consists of an array of either of these in a multi-slit assembly. Equal dimensions were chosen for the two setups with source-to-image distances (SID) of 750 mm, source-to-collimator distances (SCD) of 650 mm, and collimator-to-object distances (COD) of 50 mm, which restricts the focal length to a maximum of SCD/4 = 162.5 mm according to the Gaussian lens formula.

To model the lens and reference geometries, a mammographic tungsten-target x-ray tube was assumed,[16] and the effect of source size on maximum power was calculated as previously described.[9] A compression plate of 3 mm PMMA was inserted in both systems. The reference system was filtered with 0.5 mm aluminum. Photon-counting silicon strip detectors were assumed, and these were simply modeled with absorption only in order to isolate the effects of the filter.

Transmitted spectra of MPL and PAL filters were calculated using previously published geometrical models, which agreed well with models based on physical optics as well as measurements.[9, 10] Epoxy lenses were assumed, and to ensure manufacturing feasibility, the prism angles were chosen to equal the experimental lenses in the cited studies. The experimental PALs were made of silicon, but similar lenses have been manufactured in polymer materials.[13] Focal lengths were adjusted with the angle between the lens halves for the MPL, and with the columnar displacement for the PAL. A maximum lens length of 40 mm constrained the aperture for a certain focal length.

A 50% glandularity breast with an embedded 300 $\mu$m calcification was assumed. The spectral quantum efficiency (SQE) was used as a figure of merit for the benefit of energy filtering;[4] SQE = (SDNR$^2$/AGD) × (AGD$_{\mathrm{mono}}$/SDNR$^2_{\mathrm{mono}}$), where SDNR is the signal-difference-to-noise ratio, AGD is the average glandular dose,[17] and subscript "mono" indicates the ideal monochromatic case, which, however, also included the detector. Scattering in the object was assumed to be rejected by the multi-slit geometry. The dose reduction of the lens system compared to the reference is $1 - \mathrm{SQE}_{\mathrm{reference}}/\mathrm{SQE}_{\mathrm{lens}}$.

The resolution in the detector-strip direction is unaffected by the lens. In the scan direction, however, the lens-filtered system has a modulation transfer function (MTF) that is given by $\mathrm{MTF}_{\mathrm{aperture}} \times \mathrm{MTF}_{\mathrm{image}} \times \mathrm{MTF}_{\mathrm{scan}} \times \mathrm{MTF}_{\mathrm{add}}$. In this cascade, $\mathrm{MTF}_{\mathrm{aperture}}$ is the contribution by the lens aperture, which is the Fourier transform of the lens transmission function, scaled by a factor COD/$s_{\mathrm{i}}$, where $s_{\mathrm{i}}$ is the lens-to-collimator distance. The transmission function is a Gaussian for an MPL,[12] and exponentially decreasing towards the periphery for a PAL.[14] $\mathrm{MTF}_{\mathrm{image}}$ is the MTF contribution from the image of the source, which is a sinc function if we assume a rectangular source, scaled by a factor $s_{\mathrm{i}}/s_{\mathrm{o}} \times (\mathrm{COD} + s_{\mathrm{i}})/s_{\mathrm{i}}$, where $s_{\mathrm{o}}$ is the source-to-lens distance. $\mathrm{MTF}_{\mathrm{scan}}$ and $\mathrm{MTF}_{\mathrm{add}}$ are the contributions from the scan step, and from misaligned detector units, which are both sinc functions, scaled by a factor (SCD + COD)/SID. The resolution of the reference system is determined correspondingly by the source, the slit, the scan step, and misaligned detector units.[11] The figure of merit to compare resolution was chosen as the spatial frequency at an MTF of 0.5 ($MTF_{0.5}$) measured at COD.

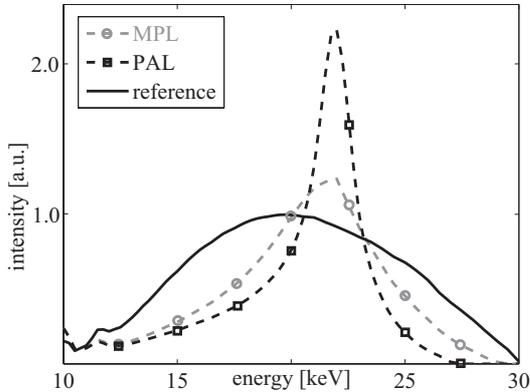

Figure 2. Typical spectra of the lens- and reference geometries.

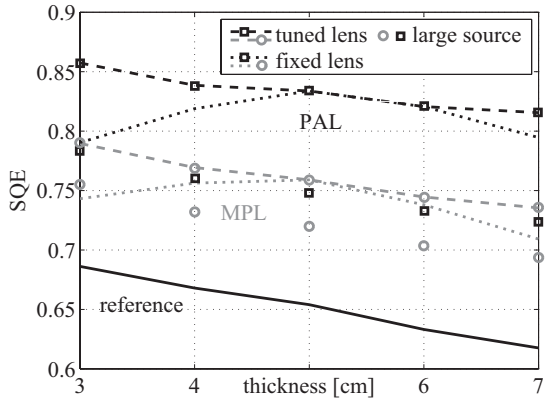

Figure 3. Spectral quantum efficiency (SQE) for the MPL (circles), PAL (squares), and reference geometries. The lens filters with all parameters tuned in each point is shown for 25 and 100 $\mu$m sources, and the fixed filters that were optimized at 50 mm are shown for the 25 $\mu$m source.

RESULTS AND DISCUSSION

Figure 2 shows typical spectra for the lens and reference systems, optimized for a 50 mm breast and with matching scan time. The peak energy was 22 keV for both lens-filtered spectra.

The SQE as a function of breast thickness is plotted in Fig. 3 for the lens systems with 25 and 100 $\mu$m sources, and for the reference system. Focal length, peak energy, acceleration voltage, and aperture of the lens systems were tuned under constraints to maximize the SQE for each breast thickness and source size, with details in Table 1. Also shown in Fig. 3 is the SQE for a fixed filter that was optimized at 50 mm breast thickness, and tuned only with the acceleration voltage. For the absorption-filtered system, the acceleration voltage was tuned in accordance with a previous optimization.[4]

The MPL filter reduced the dose compared to the reference system 13 – 16% at matching scan times and an improved spatial resolution. Note that the resolution is almost constant when going to a larger source, but a demagnifying setup ($s_i < s_o$) maximizes the SQE. The improved spatial resolution of the MPL-filtered system is not necessarily desirable because the reference system may already be good enough, but it is not possible to trade resolution for SQE or flux because the aperture of the MPL is limited. The PAL filter allows for larger apertures and both scan time and resolution can be matched to the reference system, resulting in a dose reduction of 20 – 24%. Interestingly, the dose of the PAL-filtered system is only $\sim 20\%$ higher than for the monochromatic case.

The SQE for the fixed lens filters decreased only slightly away from the optimum, which shows that the performance is almost constant even if the filter cannot be changed or tuned in practice. When going from a 25 to a 100 $\mu$m source, the reduction in SQE was also not large, but the benefit compared to the reference system decreased 30 – 40% across the breast thickness range, and it is evident that a thin line-shaped source is more or less required. It has previously been concluded that bright-enough micro-focal sources may be feasible,[9] but this issue needs further attention.

CONCLUSIONS

A theoretical model was used to investigate and compare the benefit of two different refractive x-ray lenses – the MPL and the PAL – for energy filtering in mammography. The dose of the PAL filter was $\sim 20\%$ lower than for an absorption filter, $\sim 10\%$ lower compared to the MPL, and only $\sim 20\%$ higher than for a monochromatic beam. An increased resolution compared to the reference system was found with the MPL filter, but the higher dose reduction of the PAL filter is likely more beneficial. On the other hand, the complexity of the PAL is higher, and it is probable that the MPL can be better optimized in the practical case. An investigation of the feasibility of small x-ray sources remains.

Table 1. Comparison of lens and reference (ref.) geometries in terms of spectral quantum efficiency (SQE) for breast thicknesses $30 - 70$ mm. Dose reduction and scan time are relative the reference system. MTF$_{0.5}$ is the spatial frequency at an MTF of 0.5. Focal length ($F$), peak energy ($E_{\text{peak}}$), acceleration voltage (kVp), and aperture were tuned under constraints to maximize the SQE for each breast thickness. A negative $F$ indicates a demagnifying setup.

|        | SQE         | dose red. [%] | scan time | MTF$_{0.5}$ [mm$^{-1}$] | kVp     | source [$\mu$m] | $F$ [mm]     | aperture [mm] | $E_{\text{peak}}$ [keV] |
|--------|-------------|---------------|-----------|-------------------------|---------|-----------------|--------------|---------------|-------------------------|
| **MPL:** | 0.79 - 0.74 | 13 - 16       | matched   | 6.3 - 6.6               | 25 - 31 | 25              | 130 - 135    | 0.23 - 0.19   | 20 - 24                 |
|        | 0.76 - 0.69 | 9 - 11        |           | 6.9 - 6.7               | 26 - 31 | 100             | −140 - 145   | 0.23 - 0.19   | 21 - 25                 |
| **PAL:** | 0.86 - 0.82 | 20 - 24       | matched   | matched                 | 24 - 29 | 25              | −160         | 0.55 - 0.60   | 19 - 23                 |
|        | 0.78 - 0.72 | 12 - 15       |           |                         | 24 - 30 | 100             | −95 - 85     | 0.23 - 0.20   | 20 - 24                 |
| **ref.:** | 0.68 - 0.62 | -             | -         | 5.5                     | 28 - 32 | 450             | -            | -             | -                       |

REFERENCES

1. Motz, J, Danos, M (1978) Image information content and patient exposure. *Med. Phys.*, **5**, 8–22.
2. Pfahler, G (1906) A roentgen filter and a universal diaphragm and protecting screen. *Trans. Am. Roentgen Ray Soc.*, 217–224.
3. Jennings, R, Eastgate, R, Siedband, M, Ergun, D (1981) Optimal x-ray spectra for screen-film mammography. *Med. Phys.*, **8**, 629–639.
4. Åslund, M, Cederström, B, Lundqvist, M, Danielsson, M (2006) Optimization of operating conditions in photon counting multi-slit mammography based on si-strip detectors. *Proc. of SPIE, Medical Imaging 2006*, **6142**.
5. Suortti, P, Thomlinson, W (2003) Medical applications of synchrotron radiation. *Phys. Med. Biol.*, **48**, R1–R35.
6. Piestrup, M, Wu, X, Kaplan, V, Uglov, S, Cremer, J, Rule, D, Fiorito, R (2001) A design of mammography units using a quasimonochromatic x-ray source. *Rev. Sci. Instrum.*, **72**, 2159–2170.
7. Baldelli, P, Taibi, A, Tuffanelli, A, Gilardoni, M, Gambaccini, M (2005) A prototype of a quasi-monochromatic system for mammography applications. *Phys. Med. Biol.*, **50**, 2225–2240.
8. Bingölbali, A, MacDonald, CA (2009) Curved crystal x-ray optics for monochromatic imaging with a clinical source. *Med. Phys.*, **36**, 1176–1183.
9. Fredenberg, E, Cederström, B, Åslund, M, Ribbing, C, Danielsson, M (2008) A tunable energy filter for medical x-ray imaging. *X-Ray Optics and Instrumentation*, **2008**, 8 pages.
10. Fredenberg, E, Cederström, B, Nillius, P, Ribbing, C, Karlsson, S, Danielsson, M (2009) A low-absorption x-ray energy filter for small-scale applications. *Opt. Express*, **17**, 11388–11398.
11. Lundqvist, M (2003) *Silicon strip detectors for scanned multi-slit x-ray imaging*. Ph.D. thesis, Royal Institute of Technology (KTH), Stockholm.
12. Cederström, B, Cahn, R, Danielsson, M, Lundqvist, M, Nygren, D (2000) Focusing hard x-rays with old LP's. *Nature*, **404**, 951.
13. Jark, W, Pérennès, F, Matteucci, M, Mancini, L, Montanari, L, Rigon, L, Tromba, G, Somogyi, A, Tucoulou, R, Bohic, S (2004) Focusing x-rays with simple arrays of prism-like structures. *J. Synchrotron Rad.*, **11**, 248–253.
14. Cederström, B, Ribbing, C, Lundqvist, M (2005) Generalized prism-array lenses for hard x-rays. *J. Synchrotron Rad.*, **12**, 340–344.
15. Cederström, B, Lundqvist, M, Ribbing, C (2002) Multi-prism x-ray lens. *Appl. Phys. Lett.*, **81**, 1399–1401.
16. Boone, J, Fewell, T, Jennings, R (1997) Molybdenum, rhodium, and tungsten anode spectral models using interpolating polynomials with application to mammography. *Med. Phys.*, **24**, 1863–74.
17. Boone, J (1999) Glandular breast dose for monoenergetic and high-energy x-ray beams: Monte Carlo assessment. *Radiology*, **203**, 23–37.